\documentclass[%
 reprint,
superscriptaddress,
 amsmath,amssymb,
 aps,
prb,
floatfix,
]{revtex4-2}

\usepackage{graphicx}
\usepackage{dcolumn}
\usepackage{bm}
\usepackage{amsmath}

\usepackage{xcolor}
\usepackage{soul}
\usepackage[toc,page]{appendix}
\usepackage{braket}
\usepackage{ dsfont }

\soulregister{\ref}7

\begin{document}

\title{Appearance of Multiple Spectral Gaps in Voltage-Biased Josephson Junctions Without Floquet Hybridization}

\author{Teng Zhang}
 \affiliation{%
 Department of Physics and Astronomy, Purdue University, West Lafayette, Indiana 47907, USA
}%
 \affiliation{Birck Nanotechnology Center, Purdue University, West Lafayette, Indiana 47907, USA}

\author{Tatiana de Picoli}
 \affiliation{%
 Department of Physics and Astronomy, Purdue University, West Lafayette, Indiana 47907, USA
}%

\author{Tyler Lindemann}
 \affiliation{%
 Department of Physics and Astronomy, Purdue University, West Lafayette, Indiana 47907, USA
}%
 \affiliation{Birck Nanotechnology Center, Purdue University, West Lafayette, Indiana 47907, USA} 
  \affiliation{Microsoft Quantum Lab West Lafayette, West Lafayette, Indiana 47907, USA}

\author{Jukka I. Väyrynen}
 \affiliation{%
 Department of Physics and Astronomy, Purdue University, West Lafayette, Indiana 47907, USA
}%
 
\author{Michael J. Manfra}
\email{mmanfra@purdue.edu}
 \affiliation{Department of Physics and Astronomy, Purdue University, West Lafayette, Indiana 47907, USA }
 \affiliation{Birck Nanotechnology Center, Purdue University, West Lafayette, Indiana 47907, USA} 
 \affiliation{Microsoft Quantum Lab West Lafayette, West Lafayette, Indiana 47907, USA}
 \affiliation{School of Materials Engineering, Purdue University, West Lafayette, Indiana 47907, USA}
 \affiliation{Elmore Family School of Electrical and Computer Engineering, Purdue University, West Lafayette, Indiana 47907, USA}

 \date{\today}
 
\begin{abstract}
A time-periodic drive enables the engineering of non-equilibrium quantum systems by hybridizing Floquet sidebands. We investigated DC voltage-biased planar Josephson junctions built upon epitaxial Al/InAs heterostructures in which the intrinsic AC Josephson effect is theoretically expected to provide a time-periodic drive leading to Floquet hybridization. Tunneling spectroscopy is performed using probes positioned at the ends of the junction to study the evolution of the local density of states. With applied drive, we observe multiple coherence peaks which are studied as a function of DC voltage bias and in-plane magnetic field. Our analysis suggests that these spectral gaps arise from a direct mesoscopic coupling between the tunneling probe and the superconducting leads rather than from a Floquet-driven gap opening. Our numerical simulations indicate that an increase in the ratio of junction width to coherence length will enhance the contribution of Floquet hybridization. This work lays a foundation for the exploration of Floquet physics utilizing voltage-biased hybrid superconductor-semiconductor Josephson junctions and provides means for distinguishing direct couplings from genuine Floquet effects.
\end{abstract}

\maketitle

\section{Introduction}
Shallow III-V semiconductor quantum wells hosting a high-mobility two-dimensional electron gas (2DEG) with strong spin-orbit coupling coupled to epitaxial aluminum realize a hybrid platform for exploration of topological superconductivity in a tunable Josephson junction geometry \cite{Pientka.2017, Fornieri.2019, Ren.2019, Dartiailh.2021, Coraiola.2023, Banerjee.2023f40q, Banerjee.2023zyr}. A key attribute of Josephson junctions is that the superconducting phase difference between the leads may be utilized to generate novel state configurations. Recently, static phase bias achieved by embedding a planar Josephson junction in a superconducting loop threaded by magnetic flux has been widely used to promote and explore topological superconductivity \cite{Fornieri.2019, Banerjee.2023f40q, Banerjee.2023zyr}. When the DC current through a planar Josephson junction exceeds the critical current, a finite DC voltage bias ($V_J$) is generated and the superconducting phase difference becomes time dependent, resulting in an intrinsic alternating current at a frequency proportional to $V_J$. This process is known as the AC Josephson effect. The AC Josephson effect is an attractive path toward dynamic control of the superconducting state. Recently, a proposal for Floquet-enhanced topological superconductivity driven by the intrinsic AC Josephson effect has been put forth \cite{peng21_floquet, Zhang.2021}. In this proposal, the system is effectively driven at the Josephson frequency $f=2eV_J/h$. Typical Josephson frequencies are tens of GHz for a planar Josephson junction built upon epitaxial Al/InAs heterostructures. This relatively high drive frequency may lead to less stringent requirements on the coherence time for the Floquet hybridization state, a crucial parameter for Floquet engineering \cite{rudner2020, Rudner.20201rh}. Although Floquet physics has attracted significant theoretical attention, experimental study in condensed matter systems is challenging and still in its infancy. It is important to develop experimental platforms and identify signatures of Floquet physics that may be used to distinguish among competing mechanisms. For example, both photon-assisted tunneling and Floquet theory may produce a similar replica of the tunneling conductance peaks under microwave drive, as reported in Refs. \cite{Park.2022, Haxell.2023gfa}.

Here we present an experimental and theoretical study of gate-tunable planar Josephson junctions built upon a hybrid epitaxial Al-InAs 2DEG heterostructure in which we measured the tunneling conductance at the two ends of the central InAs strip using probes formed from quantum point contacts (QPCs), as shown in Fig. \ref{set_up}. We study the evolution of the tunneling density of states (DOS) at the ends of the junction as a function of DC bias current and the applied in-plane magnetic field. When the DC current generates a finite voltage across the Josephson junction, we observe the emergence of two pairs of conductance peaks. One pair of conductance peaks is centered around zero bias, while the other pair is centered at the bias voltage $V_{J}$. With an increasing DC bias current or an increasing in-plane magnetic field, the conductance peaks disperse, and the voltage separation between the coherence peaks is reduced. Comparing our experimental results with predictions of Floquet theory and a simple three-terminal model of our device, which assumes the tunnel probe couples directly to the two superconducting leads, we attribute our experimental observations to strong coupling between the normal tunneling probe and the two proximity-induced superconducting InAs leads. This direct coupling in our mesoscopic geometry is the dominant contribution when the superconducting coherence length is longer than the physical separation of the tunnel probes and superconducting leads. Further analysis indicates that increasing the ratio of the Josephson junction width to the superconducting coherence length will promote Floquet hybridization over direct coupling in our platform. We provide estimates of the necessary geometry and assess the feasibility of future improvement. This study will guide future attempts to generate Floquet topological superconductivity in hybrid superconductor-semiconductor systems.

\section{Device Fabrication and Measurement}

\begin{figure}
\includegraphics[width=0.5\textwidth]{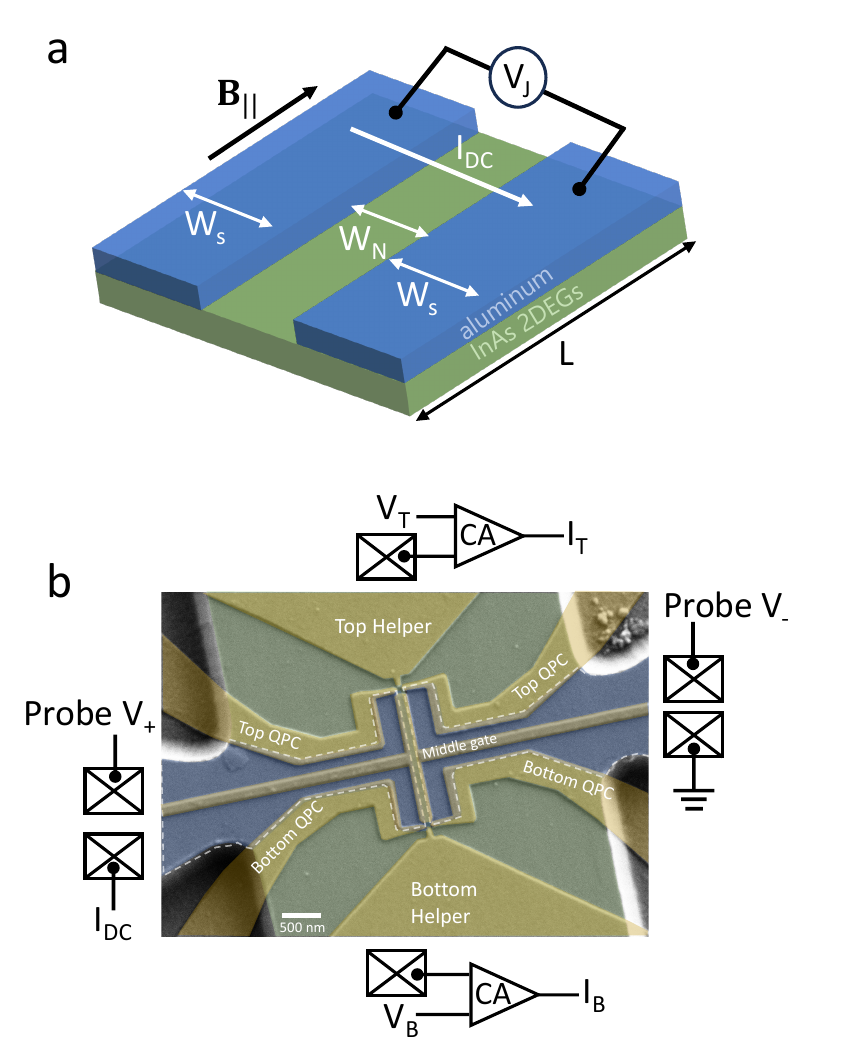}
\caption{\label{set_up} (a) Schematic of a planar Josephson junction fabricated on an epitaxial Al/InAs heterostructure. The epitaxial aluminum layer is shown in blue, while the InAs 2DEG is shown in green. DC current $I_{DC}$ is applied across the device, producing a voltage $V_J$ when the critical current is exceeded. Two pairs of QPCs (not shown) are located at the ends of the central bare InAs region to form tunnel junctions to probe the local density of states. The direction of the in-plane magnetic field $B_{\parallel}$ is indicated with an arrow. (b) False-color scanning electron micrograph of a typical device with a schematic of the measurement circuit. The DC voltage bias $V_{T(B)}$ on the tunneling probe is referenced to the ground. The Al layer is wet-etched in the green-shaded region and is left untouched in the blue-shaded region. The white dashed lines indicate the boundary of the Al pattern. All top gates (yellow-shaded regions) are separated from the heterostructure by an 18~nm hafnium oxide layer.}
\end{figure}

The heterostructure used in this study was grown by molecular beam epitaxy on an InP substrate. The 2DEG resides in the InAs quantum well separated from the epitaxial Al layer by a 10 nm In$_{0.75}$Ga$_{0.25}$As top barrier. This material system has been systematically characterized and is described in Ref. \cite{Zhang.2023}. Mesas for Josephson junctions are defined using a dilute phosphoric acid and citric acid solution. After mesa definition, the planar Josephson junctions are patterned by selectively wet etching the Al layer with Transene D at 50 $^\circ$C for 9 seconds. An 18~nm hafnium oxide layer was deposited globally by atomic layer deposition. Finally, gate electrodes are fabricated by the deposition of a Ti/Au metal stack and a standard lift-off process. The laterally adjacent Al surfaces serve as ohmic contacts to the 2DEG. For the device studied here, the width of the Al leads ($W_S$) is 300~nm, the length ($L$) is 1.6 $\mu$m, and the width of the bare InAs region ($W_N$) is 100~nm as shown in Fig.~\ref{set_up}(a) and (b). 

The transmission of the QPCs is tuned to set the high-bias conductance to $0.015 G_0$, where $G_0=e^2/2h$. Since the top and bottom QPC gate voltages are set more negatively than the 2DEG depletion voltage, the QPC gates also deplete electrons adjacent to the patterned aluminum, helping to define the conduction path through the Josephson junction. The middle gate voltage ($V_{middle}$) controls the electron density in the uncovered InAs region. For the heterostructure used in this study, the Hall bar measurements show a maximum mobility $\mu_{peak} = 57,000$ cm$^2$/Vs at a 2DEG density $n_{2DEG} = 0.6 \times 10^{12}$ cm$^{-2}$ at T=10~mK. The electronic mean free path extracted at peak mobility is 750~nm. Andreev reflections at the interface between the Al film and InAs 2DEG beneath it introduce particle and hole correlations inside InAs 2DEG, leading to an induced superconducting gap. This proximity-induced hard superconducting gap ($\Delta_{ind}$) is approximately 200 $\mu$eV and is extracted from tunneling data measured in a nearby superconductor-QPC-semiconductor junction. The interface between the Al film and InAs 2DEG beneath is highly transparent, as evidenced by the hard induced gap (see Fig.~\ref{Induced gap} and Ref.~\cite{Zhang.2023}). The superconducting coherence length is estimated to be $\xi_s = \hbar v_f/\pi\Delta_{ind} = 730$~nm, where $v_f$ is the Fermi velocity in InAs 2DEG at $n_{2DEG} = 0.6 \times 10^{12}$ cm$^{-2}$. Since the width of the Josephson junction (100~nm) is smaller than both the mean free path and the superconducting coherence length, our devices are in the short, ballistic regime. The transparency of the interface between the bare semiconductor region and the Al-covered region is close to unity, as shown in the Supplementary Materials and as reported in the literature in similar heterostructures grown in our laboratory~\cite{Fornieri.2019}.

The circuit for our transport measurements is depicted in Fig.~\ref{set_up}(b). DC current ($I_{DC}$) is applied to the lower left ohmic contact and drains to the lower right ohmic contact. Once $I_{DC}$ exceeds the critical current of the Josephson junction, a DC voltage $V_J$ is generated between the two superconducting leads. The potential difference across the junction is measured at the upper left and right ohmic contacts such that $V_{J} = V_+ - V_-$. Low-frequency ($<$100 Hz) AC plus DC voltage bias $V_{T(B)}$ is applied to the top (bottom) ohmic contacts through current amplifiers, and the currents at the top (bottom) ohmic contacts are measured using a standard lock-in measurement technique, with conductance of the top (bottom) tunnel junction given by $G_{T(B)} = \frac{dI_{T(B)}}{dV_{T(B)}}$. Tunneling spectroscopy is performed by measuring conductance as a function of $V_{T(B)}$. The tunneling conductance is proportional to the local DOS. All transport measurements are conducted in an Oxford Triton 500 cryogen-free dilution refrigerator with a base mixing chamber temperature of 10~mK equipped with a 6-1-1~T vector magnet.

\section{Experimental Results and Analysis}
\begin{figure*}
\includegraphics[width=1\textwidth]{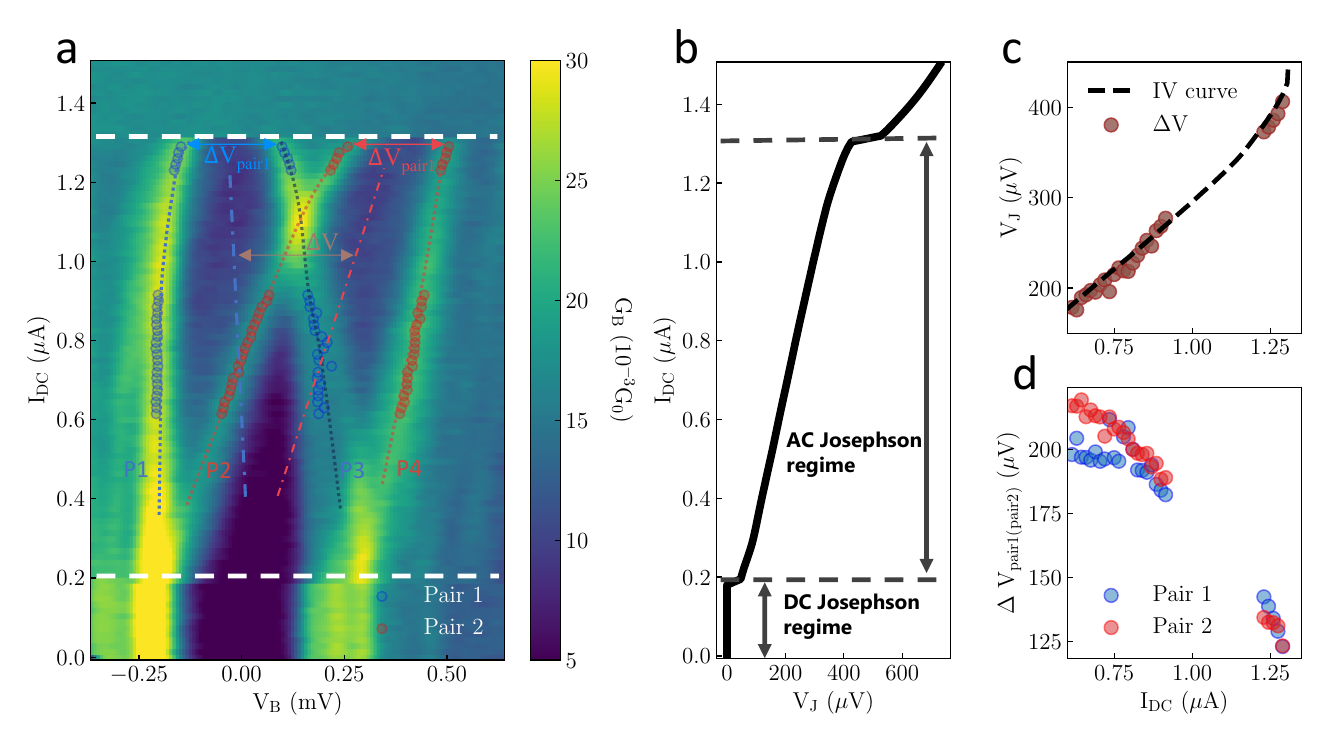}
\caption{\label{main} (a) Local tunneling conductance measured at the bottom probe as a function of $V_B$ and $I_{DC}$ at $B_{\parallel} = 100$~mT. The region between two white dashed lines indicates the AC Josephson regime. As a guide to the eye, the four peaks are highlighted by red and light blue dotted lines, and the centerpoints of these pairs are highlighted by blue (red) dashed lines. Light blue (red) lines track P1 and P3 (P2 and P4). The centerpoints of these pairs of peaks are expected to be at $V_B \approx 0$ ($V_B \approx V_J$) in the AC Josephson regime. Red and blue open circles mark the positions used to calculate $\Delta V$, as indicated with the light brown arrow, and $\Delta V_{pair1(pair\,2)}$, as labeled with blue (red) arrows and plotted in (c) and (d). (b) The I-V data at $B_{\parallel} = 100$~mT is displayed. The DC and AC Josephson effect regimes are delineated by black dashed lines. (c) Brown dots represent the voltage difference between the centers of two pairs of conductance peaks, $\Delta V$. The black dashed line indicates I-V data. (d) Red (blue) dots show $\Delta V_{pair\,1(pair\,2)}$ extracted from the pair of conductance peaks centering at $V_{B} \approx 0$ ($V_{B} \approx V_J$). Panels (c) and (d) share the same x-axis.}

\end{figure*}

Fig.~\ref{main}(a) displays tunneling spectroscopy measured at the bottom tunneling probe as a function of $I_{DC}$ and $V_B$ at $B_{\parallel} = 100~$mT and $V_{middle} = -0.75~$V. Fig.~\ref{main}(b) presents I-V characteristics measured simultaneously with the tunneling spectra. As indicated by two white dashed lines in Fig.~\ref{main}(a) and two black dashed lines in Fig.~\ref{main}(b), the entire investigated range of $I_{DC}$ can be divided into three regimes: (i) the DC Josephson regime, (ii) the AC Josephson regime, and (iii) the normal regime where $eV_J > 2\Delta_{Al}$ such that superconductivity is destroyed. The DC Josephson regime is defined by $I_{DC}\leq I_c$ where $I_{c} = 0.2~\mu$A in our device. In this regime, the supercurrent is carried by dissipationless Andreev bound states (ABSs) located in the bare 2DEG region confined from both sides by Al-covered InAs and may be associated with the $V_J=0$ regime in the I-V curve, as shown in Fig.~\ref{main}(b). In the AC Josephson regime, $I_{DC} > I_{c}$ such that the DC voltage $V_{J}$ is non-zero but smaller than twice the superconducting gap of the two Al leads, $\Delta_{Al}$. In this region, the supercurrent is time-dependent, oscillating with frequency of $2eV_J/\hbar$, where $e$ is the charge of the electron, and $\hbar$ is the Planck constant. This AC Josephson regime is the main focus of our study. 

Without DC current bias and at small in-plane magnetic field, the superconducting phase difference between the two leads $\Delta \phi$ is close to zero, resulting in the minimum energy of the Andreev bound states residing close to $\Delta_{ind}$, the proximity-induced superconducting gap. As shown in Fig.~\ref{main}(a), at $I_{DC} = 0~\mu$A, the minimum energy of the ABS is extracted by examination of the tunneling spectra and is estimated to be $200~\mu$eV, approximately equal to $\Delta_{ind}$ for this heterostructure. Measurements of $\Delta_{ind}$ for this wafer are discussed in the Supplementary Materials. Each Andreev bound state carries a supercurrent of $-\frac{2e}{h}\frac{dE_{ABS}}{d\phi}$, where $E_{ABS}$ is the energy of the ABS. As the DC current bias increases, the system modifies $\Delta \phi$ to match the supercurrent with the DC current bias, leading to the modulation of $E_{ABS}$. However, in the DC Josephson regime, modulation of $E_{ABS}$ is not readily identified in the tunneling spectra shown in Fig.~\ref{main}(a). 

A likely reason for absence of modulation in $E_{ABS}$ is that the tunneling probe is not coupling to the Andreev bound states but instead directly to the proximitized superconducting leads surrounding the junction. We note that the critical current may be reduced in our geometry due to varying transmissions of individual ABS \cite{Kjaergaard.2017, Nichele.2020, Schäpers.1998r3g}. The injection of electrons from the normal tunnel probe may also decrease the critical current. A reduction in critical current may also contribute to the absence of $E_{ABS}$ modulation. A more detailed discussion of critical current suppression can be found in the Supplementary Materials.

When $I_{DC} > I_{c}$, a finite DC voltage $V_J$ is generated as seen in the AC Josephson regime of the I-V data presented in Fig.~\ref{main}(b). In this regime, four conductance peaks (P1-P4) become visible and disperse with an increase of $I_{DC}$, as illustrated by the red and blue dotted lines in Fig.~\ref{main}(a). The midpoint between the conductance peaks P1 and P3 is centered around zero bias, while the midpoint between the peaks P2 and P4 disperses to a higher voltage as $I_{DC}$ increases, as highlighted in blue for P1, P3 and red for P2, P4 in Fig.~\ref{main}(a). The separation in voltage between the conductance peaks decreases as $I_{DC}$ increases, as shown in blue for P1, P3 and in red for P2, P4 in Fig.~\ref{main}(a). 

When $eV_J > 2\Delta_{Al}$, or $I_{DC} > 1.3~\mu$A in Fig.~\ref{main}(a), the superconductivity in the Al film is destroyed, eliminating all sharp features in the tunneling spectra. This behavior emphasizes that the four conductance peaks observed in the AC Josephson regime originate from superconductivity. The superconducting gap of the Al film is estimated to be $250~\mu$eV from $V_J$ at the transition current $I_{DC} = 1.3~\mu$A.

In Fig.~\ref{main}(a), the blue and red points track the dispersion of the conductance peaks (P1 - P4) with increasing $I_{DC}$. Points are identified by a peak finding script in the range $I_{DC} = 0.6\,\mu$A to $I_{DC} = 1.3\,\mu $A. In the AC Josephson regime, when $I_{DC} \leq 0.6\,\mu $A, P3, and P4 overlap such that they cannot be reliably distinguished. Similarly, from $0.92~\mu$A $\leq I_{DC} \leq 1.23~\mu$A, P2 and P3 merge and cross. The voltage separation between the centers of two pairs of conductance peaks is defined as follows:
\begin{equation}
    \Delta V = \frac{V_2+V_4}{2}-\frac{V_1+V_3}{2} 
\end{equation}
where $V_i$ indicates the voltage of $i^{th}$ conductance peak as shown in Fig.~\ref{main}(a). $\Delta V$ is proportional to the energy separation between the two apparent gaps. In Fig.~\ref{main}(c), $\Delta V$ as a function of $I_{DC}$ is plotted with brown circles and overlaid on a plot of $V_J$ as a function of $I_{DC}$. It is apparent that $\Delta V$ tracks $V_{J}$ in the studied $I_{DC}$ range.

Another quantity extracted from the data in Fig.~\ref{main}(a) is the voltage difference between the two conductance peaks for each pair, which  we defined as:
\begin{equation}
    \begin{split}
            \Delta V_{pair\,1} = \frac{V_3-V_1}{2} \\
            \Delta V_{pair\,2} = \frac{V_4-V_2}{2}
    \end{split}
\end{equation}
$\Delta V_{pair\,1(pair\, 2)}$ defines two apparent spectral gaps. In Fig.~\ref{main}(d), $\Delta V_{pair\,1(pair\, 2)}$ are plotted as a function of $I_{DC}$ with blue and red points, respectively. In the studied $I_{DC}$ range, both $\Delta V_{pair\,1}$ and $\Delta V_{pair\,2}$ follow a similar trend as a function of $I_{DC}$: $\Delta V_{pair\,1(pair\, 2)}$ decreases from approximately 210 $\mu V$ at $I_{DC}$ = 0.6~$\mu A$ to 120 $\mu~V$ at $I_{DC}$ = 1.3 $\mu A$. As $I_{DC}$ increases, the proximity-induced superconducting gap decreases.

\begin{figure}
\includegraphics[width=0.5\textwidth]{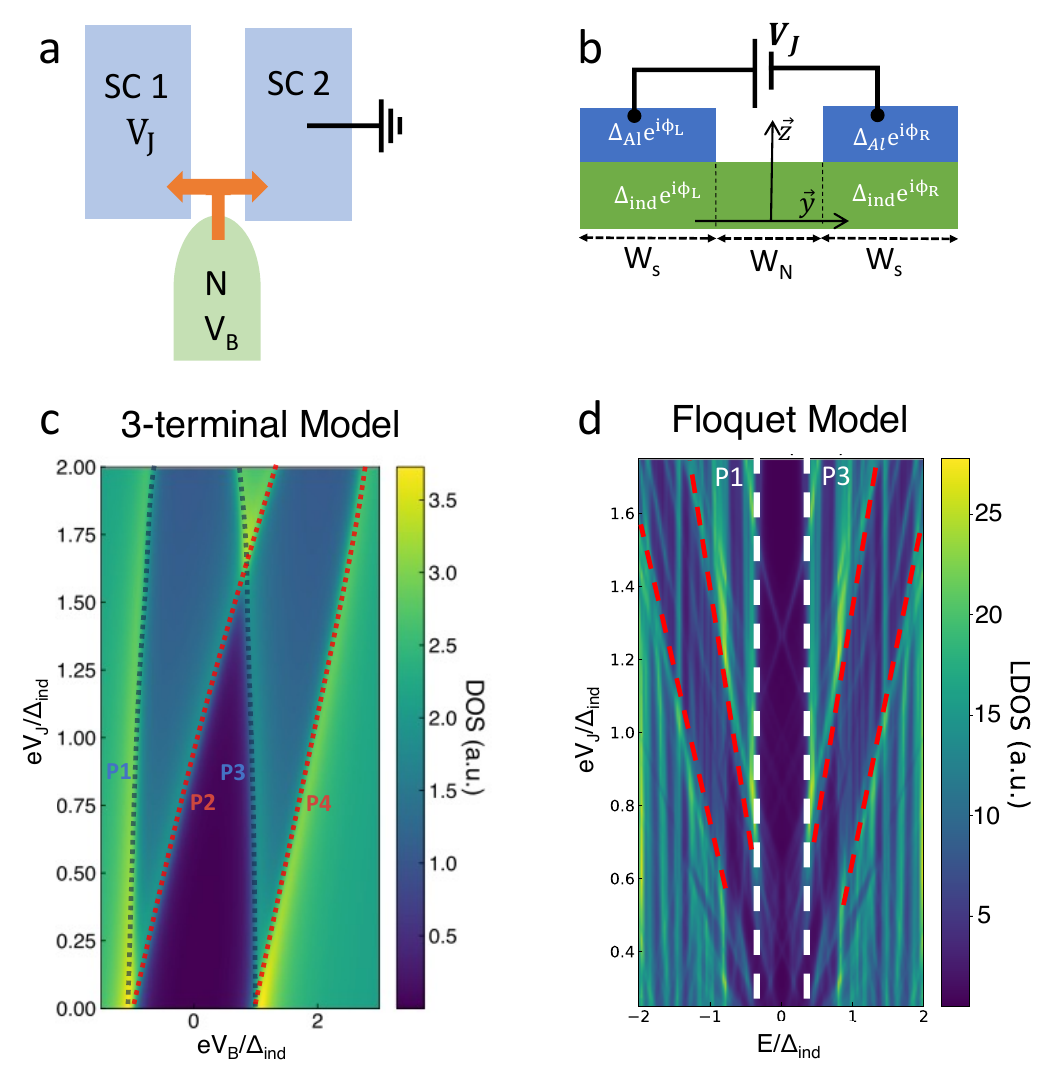}
\caption{\label{models} (a) Schematic of the 3-terminal model in which two blue squares indicate two superconducting (SC) leads connected to the green region representing the normal tunnel junction. The potential of the left SC lead is $V_J$, and the right SC lead is grounded. The normal tunnel probe is indicated by a green tip at the bottom. Orange arrows indicate the tunneling current path, which is related to the DOS of each SC lead. (b) A cross-sectional view of the Josephson junction with the superconducting pairing terms identified in each region. The coordinates used in the numerical simulation are labeled. y = 0 is located at the center of the junction. (c) DOS as a function of $V_J$ and the DC voltage $V_B$ on the tunnel probe. Red and blue dashed lines are guides to the eye, tracking two pairs of coherence peaks associated with the SC leads. (d) DOS as a function $V_J$ and energy of tunneling electrons calculated from the numerical simulation using Floquet theory. White dashed lines indicate coherence peaks centered around zero energy. On both positive and negative energy sides, low DOS regions due to Floquet hybridization are highlighted by two red dashed lines on each side.}
\end{figure}

Although multiple spectra gaps in the regime $0 < eV_J < 2 \Delta_{Al}$ are expected from Floquet hybridization, we first consider a simpler mechanism. Given that the Josephson junction under study is in the short and ballistic regime, we consider a coupling mechanism in which the normal tunnel probe is directly coupled to the two superconducting (SC) leads positioned less than a superconducting coherence length away from the tunnel probe. In this scenario, tunneling spectroscopy probes the density of states of the region consisting of the bare InAs and two Al-covered InAs regions, which form the two SC leads of the Josephson junction. This strong coupling mechanism can be taken into account in a 3-terminal model, as shown in Fig.~\ref{models}(a). We consider a normal (N) lead at voltage $V_B$ connected to two superconductors: one superconductor (SC1) at voltage $V_J$, and the other lead (SC2) is grounded. When an electron tunnels into the junction, before changing its energy, it can enter either of the two SC leads if there are available states, as indicated by the orange arrow in Fig.~\ref{models}(a).

We may express the differential conductance $d I_N / d V_B$  as two contributions from N to SC1 and SC2. Each of the NS interfaces is described by a low-temperature tunneling conductance $dI_{NS}/dV\propto N_{BCS}(e|V|)$ \cite{tinkham2004introduction}, where $N_{BCS}(e|V|)$ is the Bardeen–Cooper–Schrieffer (BCS) DOS of proximitized InAs 2DEG.
The total differential conductance will be proportional to the sum of contributions from each NS interface. Additionally, we consider the gap suppression due to the supercurrent and Zeeman field, as shown in the supplementary materials section. 
In Fig.~\ref{models}(c), we plot the differential conductance as a function of $V_J$ and the voltage applied to the normal tunnel contact, $V_B$. In this calculation, two pairs of conductance peaks are observed, with each pair of conductance peaks corresponding to the two coherence peaks associated with each SC lead. Since SC terminal 2 is grounded, two coherence peaks (blue dotted lines in Fig.~\ref{models}(c)) of this terminal are centered around zero energy while the two coherence peaks (red dotted lines in Fig.~\ref{models}(c)) of SC1 disperse with $V_J$. This simple model quantitatively captures the main features seen in our data plotted in Fig.~\ref{main}(c). (We note that the Tien-Gordon theory~\cite{TienGordon1963} predicts additional multiphoton peaks, which we do not observe.) In addition, the suppression of the gap due to the increase in supercurrent is qualitatively consistent with the observations shown in Fig.~\ref{main}(d).

Given the presence of the AC Josephson effect, one may attempt to understand the appearance of four conductance peaks in the regime $0 < eV_J < 2 \Delta_{Al}$ in the framework of Floquet theory. For a gapped system driven at an angular frequency $\omega$, an additional Floquet gap arising from the avoided crossing between the Floquet sidebands is expected to appear at an energy given by half the drive frequency $\hbar \omega/2$ \cite{peng18_floquet, Rudner.20201rh, rudner2020, peng21_floquet, Zhang.2021}. The size of the Floquet gap is related to the strength of the interaction between the Floquet sidebands \cite{peng18_floquet, Rudner.20201rh, rudner2020, peng21_floquet, Zhang.2021}. In a Josephson junction with a time-periodic pairing potential, the size of the Floquet gap is proportional to the size of the proximity-induced superconducting gap, which determines the interaction between hole-like Floquet sidebands and electron-like Floquet sidebands. The Floquet gap will be heralded by the appearance of additional coherence peaks~\cite{rudner2020}.
The results showed in  Fig.~\ref{main}(a) for the energy separation between two apparent gaps and in Fig.~\ref{main}(d) for the decrease of $\Delta V_{pair\,1(pair\, 2)}$, are qualitatively consistent with the Floquet theory \cite{rudner2020, Rudner.20201rh, Zhang.2021, peng21_floquet}.

To determine if the four conductance peaks observed when a finite DC voltage is generated at the Josephson junction are consistent with the Floquet prediction for a driven SNS system, we perform a numerical simulation of our device using combined tight-binding~\cite{Groth_2014} and the Floquet theory. In this simulation, we assume that the superconducting phase of the left superconducting lead, $\phi_{L}$, is zero, while the phase for the right side is $\phi_{R} = 2eV_Jt/\hbar$, as illustrated in the cross section in Fig. \ref{models}(b). We then construct the Floquet Hamiltonian to obtain the DOS as a function of $V_J$ and the energy of the tunneling electron \cite{peng21_floquet,rudner2020, Rudner.20201rh, Oka.2018}, as shown in Fig. \ref{models}(d). More details of the simulation can be found in the supplementary materials section. 

The predictions of Floquet theory shown in Fig.~\ref{models} (d) differ significantly from the experimental data in Fig.~\ref{main} (a). Although the simulation of Floquet theory successfully generates the primary spectral gap and the conductance peaks centered around zero energy, qualitatively consistent with P1 and P3 in Fig.~\ref{main}(a), the observed behavior of the second pair of conductance peaks, P2 and P4, is not captured by the numerical simulation. Instead, the Floquet model generates low DOS regions due to partial Floquet hybridization that disperses to negative and positive energy as a function of increasing $V_J$. These regions are highlighted in Fig.~\ref{models} (d) by red dashed lines. Additionally, we expect the DOS to be symmetric around zero energy due to the particle-hole symmetry of the model, whereas the experimental data is asymmetric. 
Although one of the figures in~\cite{peng21_floquet} indicates asymmetric behavior, symmetric behavior is expected based on the symmetry of our device and the Floquet formulation, as reported in~\cite{rudner2020, Rudner.20201rh, Oka.2018,Liu.2019}. Therefore, we conclude that our experimental observations are not fully consistent with the behavior predicted by Floquet theory. 

\begin{figure}
\includegraphics[width=0.5\textwidth]{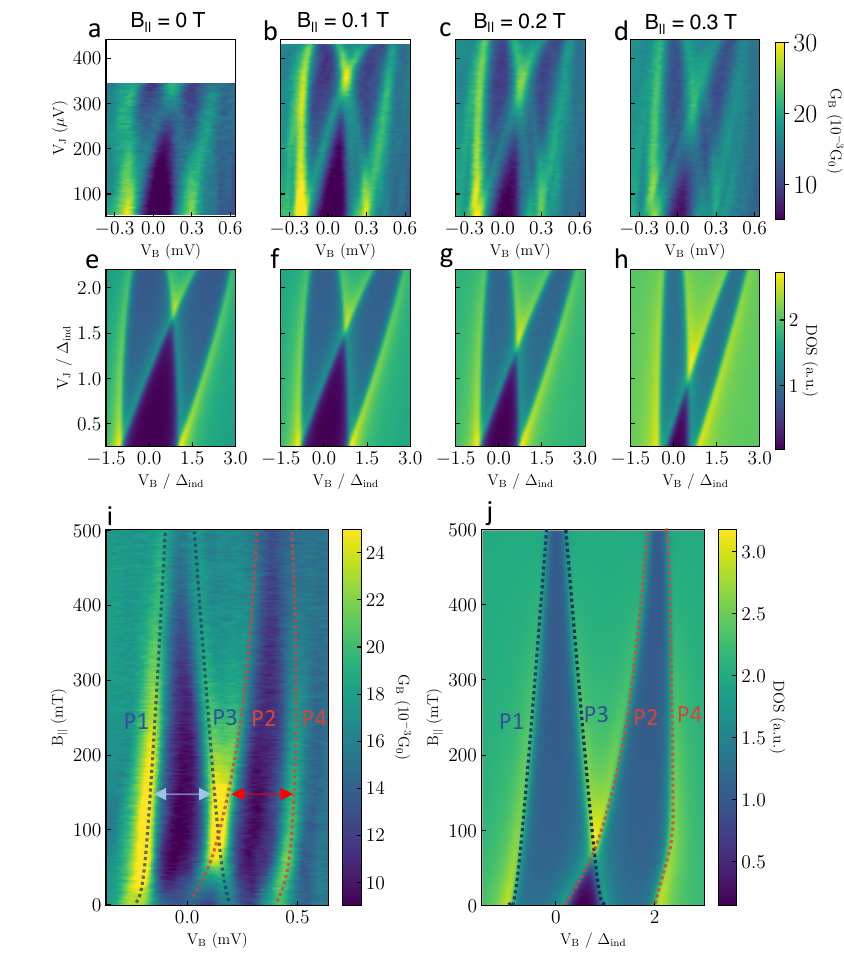}
\caption{\label{in-plane} (a) - (d): Tunneling spectra as a function of $V_J$ at $B_{\parallel} = 0$ mT, $100$ mT, $200$ mT, and $300$ mT. (a)-(d) share the same color scale. (e) - (h): Density of states (DOS) as a function of $V_J$ and $V_B$ calculated from the 3-terminal model at $B_{\parallel} = 0$ mT, $100$ mT, $200$ mT, and $300$ mT. (e)-(h) share the same color scale. (i) At $I_{DC} = 1.1\,\mu $A, tunneling spectroscopy as a function of the in-plane field, $B_{\parallel}$. As a guide to the eye, blue (red) dotted lines show two conductance peaks P1 and P3 (P2 and P4), whose centers are at $V_B \approx 0$ ($V_B \approx V_J$), and blue (red) arrow shows $\Delta V_{pair\,1}$ ($\Delta V_{pair\,2}$).  (j) At $I_{DC} = 1.1\,\mu$A, DOS as a function of $V_B$ and $B_{\parallel}$ calculated via the 3-terminal model. Similar to (i), the four peaks in the DOS are labeled with blue and red dashed lines.}
\end{figure}

In Fig.~\ref{in-plane}, we examine the density of states as a function of the in-plane magnetic field $B_{\parallel}$. Fig.~\ref{in-plane}(a)-(d) display the tunneling spectra as a function of $V_J$ in $B_{\parallel} = 0$ mT, $100$ mT, $200$ mT and $300$ mT in the AC Josephson regime. In Fig.~\ref{in-plane}(a), when $V_J \geq 350~\mu V$, $I_{DC}$ is larger than the critical current of the Al film and drives the whole system normal, while the Josephson junction at $B_{\parallel} = 100$ mT, $200$ mT, and $300$ mT is driven to normal via a larger DC voltage ($\geq 2\Delta_{Al}$) as mentioned previously. As the in-plane magnetic field increases, $\Delta V_{pair\,1(pair\, 2)}$ decreases as shown in Fig.~\ref{in-plane}(a)-(d). Figs.~\ref{in-plane}(e)-(h) show DOS as a function of $V_J$ and $V_B$ calculated with the aforementioned 3-terminal model at $B_{\parallel} = 0$ mT, $100$ mT, $200$ mT and $300$ mT. The predictions of the 3-terminal model match the experimental data for the corresponding in-plane magnetic fields; therefore, the decrease in $\Delta V_{pair\,1(pair\, 2)}$ is due to the closing of the superconducting gap on the SC leads originating from the Zeeman energy.

At a constant DC current $I_{DC} = 1.1~\mu$A, the tunneling conductance as a function of the in-plane magnetic field is shown in Fig.~\ref{in-plane}(i). As the in-plane field and the Zeeman energy increase, the size of both the first gap bounded by P1 and P3 and the second gap bounded by P2 and P4 decreases, as tracked by the decrease of $\Delta V_{pair\,1}$ and $\Delta V_{pair\,2}$ in Fig.~\ref{in-plane}(i). The spectral gap bounded by P2 and P4 shifts to the more positive voltage, whereas the first gap is anchored around zero energy. This feature is due to $V_J$ increasing as $B_{\parallel}$ increases, as shown in the supplementary material. Tunneling DOS as a function of $V_J$ and $V_B$ at $I_{DC} = 1.1\,\mu$A calculated from the 3-terminal model is shown in Fig.~\ref{in-plane}(j) with blue and red dashed lines to label P1-P4. Fig.~\ref{in-plane}(j) captures the key characteristics of the experimental data, further indicating that the spectral features originate from direct tunneling into the two proximitized InAs regions. 

The effective Landau {\it g}-factor extracted from Fig.~\ref{in-plane}(i) is approximately 8, which is higher than a previously reported value for proximitized InAs 2DEG~\cite{Nichele.2017}. We attribute this difference to the destruction of induced superconductivity by the orbital effect of the in-plane magnetic field~\cite{Pientka.2017, Fornieri.2019, Banerjee.2023zyr, Banerjee.2023f40q}. Indeed, the cross-sectional dimensions of our system of width $2W_S + W_N = 700$~nm and thickness 10~nm, with observed critical field $B_C = 650$~mT, are comparable to a previous study of a similar heterostructure in which suppression of superconductivity was attributed to the orbital effect~\cite{Banerjee.2023zyr}. 

\begin{figure}
\includegraphics[width=0.5\textwidth]{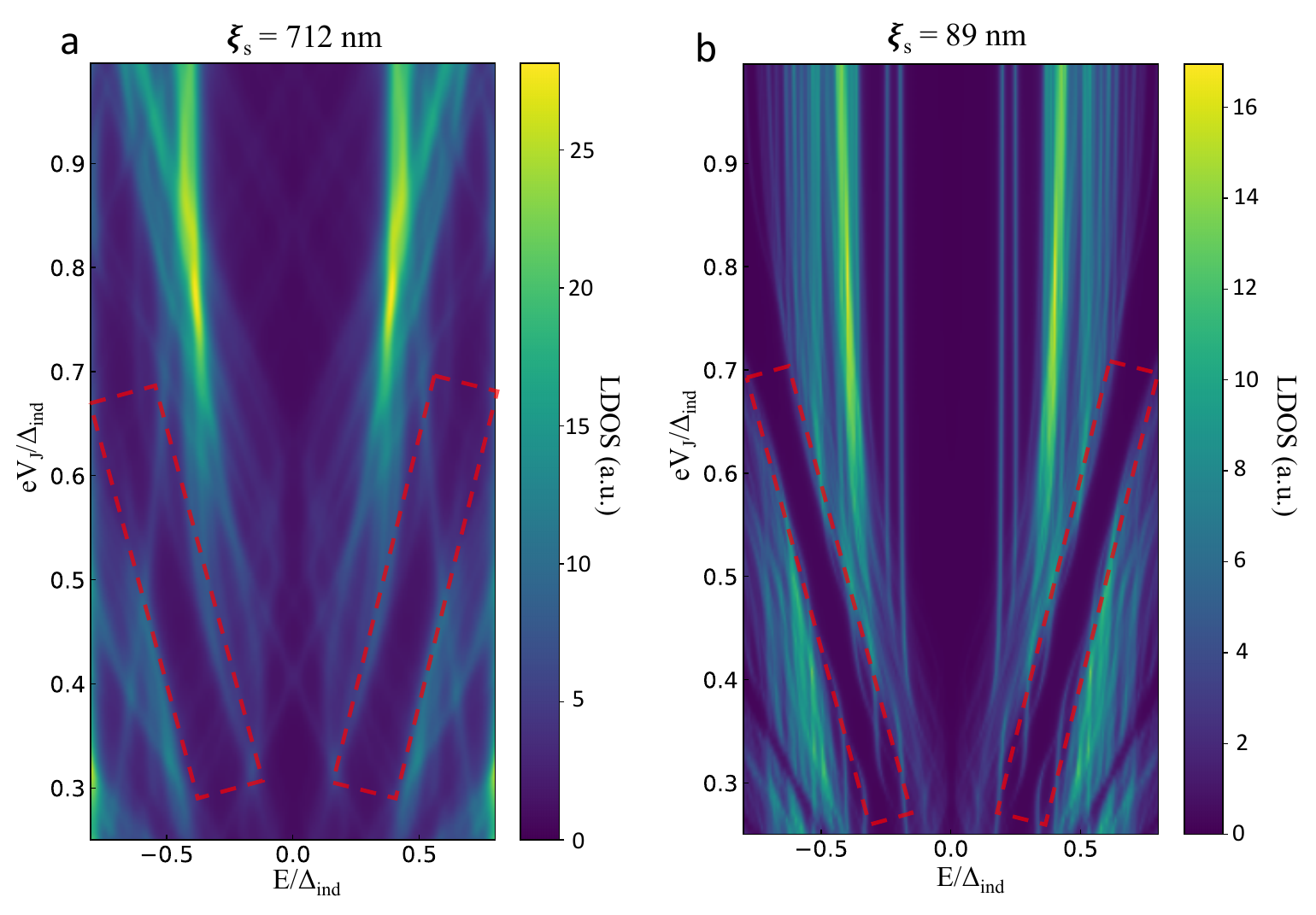}
\caption{\label{xi_study} (a) Higher detail plot of simulation data shown in Fig. \ref{models}(d). Here, the superconducting coherence length is set to $\xi_s = 712$~nm. (b) Numerically simulated local density of states (LDOS) as a function of $V_J$ and energy of the tunneling electron, with $\xi_s = 89$~nm. All other parameters are the same as the simulation in Fig. 5(a). The simulated junction width is $W_N = 100$~nm. }
\end{figure}

Given that our experimental observations can be explained without incorporating Floquet sidebands, we may ask under which conditions the Floquet sidebands would be more visible. Additional numerical simulations using Floquet theory with different superconducting coherence lengths $\xi_s$ provide some insight. Two examples with $\xi_s$=712~nm and  $\xi_s$=89~nm are shown in Fig. \ref{xi_study}. Fig.~\ref{xi_study}(a) displays details of the simulation data of Fig.~\ref{models}(c) with $\xi_s$=712~nm, close to the estimated value for our system in the clean limit. Fig.~\ref{xi_study} (b) shows the local DOS (LDOS) as a function of $V_J$ and the energy of the tunneling electron with smaller coherence length $\xi_s$=89~nm,  comparable to the width of the weak link InAs, $W_N$. In Fig.~\ref{xi_study}(a) with large $\xi_s$, the Floquet gap due to the hybridization between different Floquet sidebands is weakly visible, as indicated by the non-zero LDOS inside the gap (red boxes in Fig.~\ref{xi_study} (a)). Some states inside the Floquet gap traverse it without hybridization. With a reduction of the coherence length to $\xi_s$ =89~nm, no obvious unhybridized states are observed inside the gap, as shown by vanishing LDOS inside the gap, see Fig.~\ref{xi_study}(b). The Floquet drive, arising from the time-dependent pairing term, mixes the electron and hole components so that the subgap states hybridize strongly. When $\xi_s \approx W_N$, the Josephson junction approaches the long junction limit, and more sub-gap states are present compared to the case of the short junction ($ W_N < \xi_s$). Therefore, a stronger Floquet hybridization may be expected. Our analysis indicates that a decrease (increase) in the superconducting coherence length (the width of InAs weak link, $W_N$) promotes the formation of a Floquet hybridization gap. 
However, it must be noted that if the junction becomes too long, the Andreev bound state level spacing may fall below the spectral resolution of the tunneling probe, preventing observation of the Floquet hybridization gap. Additionally, the necessary coherent coupling between Andreev bound states and the drive~\cite{Haxell.2023gfa} (assumed in our numerical simulations) may become harder to reach in a long junction due to a larger number of relaxation channels for Andreev bound states. 

\section{Discussion and Conclusions}
In summary, we have investigated the evolution of the density of states in a current-driven planar Josephson junction using tunneling spectroscopy performed with a normal lead at the end of a junction fabricated on a gate-tunable epi-Al/InAs 2DEG heterostructure. We studied the dispersion of four conductance peaks as a function of the Josephson voltage $V_J$ generated across the Josephson junction by an applied DC current and as a function of the in-plane magnetic field $B_{\parallel}$. Our experimental results are not fully consistent with a numerical simulation based on Floquet theory. We therefore compare our results with a simple 3-terminal model that assumes that the normal tunnel probe couples directly to the two superconducting leads. The tunneling conductance
calculated from this model matches our tunneling data well. Therefore, we attribute our experimental observations to a strong coupling between the tunnel probe and the two SC leads located within the superconducting coherence length from each other in our experimental geometry. To further explore Floquet physics with voltage-biased Josephson junctions in hybrid Al/InAs heterostructures, our simulations suggest increasing the width of the bare InAs region, $W_N$, to be larger than the superconducting coherence length in the proximitized region. In this case, features associated with Floquet physics should be visible within the gap region predicted by the three-terminal model. This region is defined between P2 and P3 in Fig.~\ref{models}(d) and is associated with the BCS gap in the SC leads, an energy scale larger than the one predicted for Floquet features. Moreover, the hybridization gap is expected to appear at low driving frequencies when the energy scale related to superconductivity is reduced \cite{rudner2020, Rudner.20201rh}. This condition may enlarge the parameter space available for Floquet engineering in hybrid structures. To do so, the width of the bare InAs, $W_N$, needs to be on the same order of magnitude as the superconducting coherence length, $\xi_s$ (long junction regime). Numerical simulation using Floquet theory also indicates that increasing the ratio $W_N/\xi_s$ to the order one will enhance the Floquet hybridization gap, as shown in Fig.~\ref{xi_study}. However, if the energy scale of the Floquet hybridization gap is too small, tunneling spectroscopy may not be able to resolve distinct features in the presence of finite temperatures. To increase the likely phase space for experimental observation, a larger BCS gap superconductor, e.g., lead, may be chosen as the parent superconductor.

Beyond exploiting the intrinsic AC Josephson effect as the drive mechanism, alternative external drives such as oscillating the top gate voltage~\cite{Liu.2019}, or shining electromagnetic radiation~\cite{Park.2022} may also be employed to introduce a time-periodic term into the junction. Compared to the AC Josephson effect, external drives may offer better control through independent tuning of frequency and amplitude. Of course, the strength of coupling is critical, as weak coupling necessitates a high-power drive that may heat the experimental environment, as reported in Ref.~\cite{Park.2022}.

\section{Acknowledgments} 
This work was primarily led and supported by the U.S. Department of Energy, Office of Science, National Quantum Information Science Research Centers, Quantum Science Center (TdP, JIV, MJM) and additionally supported by Microsoft Quantum (TZ, TL, MJM).

\bibliography{ref}

\pagebreak

\begin{center}
\textbf{\large Supplementary Materials}
\end{center}

\section{Floquet theory applied to SNS junction}
\label{appendix:floquet}

To simulate the Josephson junction studied experimentally, we consider an SNS junction of the proximitized InAs two-dimensional electron gas (2DEG) as shown in Fig.~\ref{set_up}(a). The system is described by the time-dependent Bogoliubov-de Gennes Hamiltonian $H(t)=\frac{1}{2}\int d\boldsymbol{r}\Psi^{\dagger}(\boldsymbol{r})\mathcal{H}(t)\Psi(\boldsymbol{r})$ written in terms of the Nambu spinor $\hat\Psi(\boldsymbol{r})=\left[\begin{array}{cccc}
\hat\psi_{\uparrow}(\boldsymbol{r}) & \hat\psi_{\downarrow}(\boldsymbol{r}) & \hat\psi_{\downarrow}^\dagger(\boldsymbol{r}) & -\hat\psi_{\uparrow}^\dagger(\boldsymbol{r})\end{array}\right]^{T}$, and with
\begin{align}
\label{eq:SNS}
    \mathcal{H}(t)=\mathcal{H}_N\tau_{z}
    +\operatorname{Re}[\Delta(t)]\tau_{x}-\operatorname{Im}[\Delta(t)]\tau_{y}, 
\end{align}
where 
\begin{equation}
    \mathcal{H}_N=-\left[\frac{\hbar^{2}\nabla^{2}}{2m^*}+\mu(y)+i\alpha\left(\sigma_{y}\partial_{x}-\sigma_{x}\partial_{y}\right)\right].
\end{equation}
The Pauli matrices $\sigma_{x,y,z}$ and $\tau_{x,y,z}$ act on the spin and particle-hole spaces, respectively. The Rashba spin-orbit coupling is represented by $\alpha$, and $m^*$ is the effective mass. The chemical potential $\mu(y)=\mu_{SC}\Theta(|y|-W_N/2)+\mu_{J}\Theta(W_N/2-|y|)$ has a value $\mu_{SC}$ in the superconducting region and $\mu_J$ inside the junction. Similarly, we write the pairing potential as $\Delta(t)=\Delta_{ind}\Theta(|y|-W_N/2)\exp[i\Theta(y)\phi(t)]$, where $\Delta_{ind}$ is the induced gap in the 2DEG and $\phi(t)=2eV_Jt/\hbar$ is the time-dependent phase difference between the two superconducting leads, determined by the Josephson voltage $V_J$ across the junction, as shown in Fig.~\ref{models}(a). 

Since the Hamiltonian is time-periodic with period $T=2\pi/\omega$ related to the Josephson frequency $\omega=2eV_J/\hbar$, we can use the Floquet theory to write the Hamiltonian in the Fourier harmonic space \cite{rudner2020}, with matrix elements given by 
\begin{equation}
\label{eq:floquet_elements}
    \mathcal{H}_{mn}^F=-m\hbar\omega\delta_{mn}+\int_0^T dt e^{i(m-n)\omega t}\mathcal{H}(t),
\end{equation}
where $m, n \in \mathds{Z} $.
Using Eq.~\eqref{eq:floquet_elements} we find that the matrix representation for the Floquet Hamiltonian in the Fourier basis labeled by $m,n$ is given by 
\begin{equation}
\label{eq:floquet_matrix}
    \mathcal{H}^{F}=\left(\begin{array}{ccc}
\ddots & V & 0\\
V^{\dagger} & H_{0}-m\hbar\omega & V\\
0 & V^{\dagger} & \ddots
\end{array}\right),
\end{equation}
with $H_0=\mathcal{H}_{N}\tau_z+\Delta_{ind}\Theta(y+W_N/2)\tau_{x}$, $V=\Delta_{ind}\Theta(W_N/2-y)(\tau_x+i\tau_y)/2$ and $V^\dagger=\Delta_{ind}\Theta(W_N/2-y)(\tau_x-i\tau_y)/2$.

To probe the time-averaged density of states in the junction, we must diagonalize the Floquet Hamiltonian numerically. By using the Floquet theory, we get the Hamiltonian of Eq.~\eqref{eq:floquet_matrix}, which is dimensionally larger than the original Hamiltonian, Eq.~\eqref{eq:SNS}, resulting in a Floquet Hamiltonian that encodes a redundancy in the quasienergies and eigenvectors. We define the first ``Floquet-Brillouin zone'' (FBZ) by the set of quasienergies in the interval $-\hbar\omega/2 <E \leq \hbar\omega / 2$~\cite{rudner2020}. Any eigenvector of Eq.~\eqref{eq:floquet_matrix} with quasienergy outside the first FBZ can be obtained from a state in the first FBZ with energy shifted by $m\hbar\omega$. To simulate the system numerically,  we will truncate the Floquet Hamiltonian to a finite number $(2M+1)$ of Fourier coefficients. This is enabled by the linear potential $-m\hbar\omega$ in Eq.~\eqref{eq:floquet_matrix}, which results in localized eigenstates in the Fourier harmonic space. The finite range $\ell_m$ of appreciable Fourier coefficients for the Floquet eigenstates can be estimated as $\ell_m\sim\mathcal{W}/(\hbar\omega)$, where $\mathcal{W}$ is the bandwidth, approximately given by the largest one of $H_0$ and $V$. The truncation will provide a good approximation for the system far from the truncation boundaries if we take  $(2M+1)\gg\ell_m$. For details, see Ref.~\cite{rudner2020}.

We simulate the planar SNS system using the \texttt{Kwant}~\cite{Groth_2014} python package, where we take a tight-binding approximation of Eq.~\eqref{eq:floquet_elements} in a finite square lattice of size $L$ by $L_y$, where $L_y=2W_S+W_N$, and truncate the Floquet matrix to the $M$-th lowest Floquet sideband, i.e., $m,n\in[-M, M]$. At this point, with the Floquet truncation and tight binding approximation, our system is still computationally costly to solve numerically for a realistic set of parameters. Considering a system of size $L\times L_y \sim  1~\mu \text{m} \times 1~\mu \text{m}$ and lattice constant of order $\sim 1~$nm, results in a  $\sim 10^6$-dimensional Hamiltonian $H_0$. Here, instead, we introduce an effective Hamiltonian approach to solve the Floquet Hamiltonian numerically. We choose an undriven SNS Hamiltonian given by $H_{SNS}=H_0+V+V^\dagger$ and solve the eigenvalue problem numerically, finding  $N$ eigenvalues closest to the Fermi level. Then, we project $H_0$, $V$, and $V^\dagger$ in Eq.~\eqref{eq:floquet_elements} on this reduced $N$-dimensional basis, which, combined with the Floquet harmonics truncation, results in a  $(2M+1)N$-dimensional Floquet Hamiltonian. Alternatively, we could have chosen another basis for the effective Hamiltonian projection, such as the SN Hamiltonian $H_0$ eigenstates. For sufficiently large $N$ (we take $N=400$), both basis choices will result in indistinguishable results for the low-energy (near Fermi level) properties.

By using the obtained eigenvectors and eigenstates, 
we calculate the time-averaged density of states (DOS) defined as~\cite{rudner2020} 
\begin{equation}
    \label{eq:DOS}
    \rho(\omega, E)=\sum_{\nu}\sum_{m=-M}^M\braket{\phi_\nu^{(m)}|\phi_\nu^{(m)}}\delta(\epsilon_\nu+m\hbar\omega-E),  
\end{equation}
where the sum over  $\nu$ is over all the eigenvalues. 
Here $\ket{\phi_\nu^{(m)}}$ is the $m$ Fourier harmonic, obtained from the eigenvector $\varphi_\nu$ corresponding to the $\nu$-th eigenvalue $\epsilon_\nu$ of Eq.~\eqref{eq:floquet_matrix},
\begin{equation}
\label{eq:floquet_eigenstates}
    \varphi_\nu=\left(\begin{array}{c}
\vdots\\
\Ket{\phi_{\nu}^{(m-1)}}\\
\Ket{\phi_{\nu}^{(m)}}\\
\Ket{\phi_{\nu}^{(m+1)}}\\
\vdots
\end{array}\right).
\end{equation}
By projecting the eigenvector to a small region on the edge of the junction with the size determined by the Fermi wavelength, we can obtain the local density of states (LDOS) in the junction. The LDOS with $\omega = 2eV_J / \hbar$  is plotted in Figs.~\ref{models}(c) and \ref{xi_study}(a) of the main text. 

For the LDOS simulation shown in this work, we take the parameters $W_S=300~\text{nm}$, $W_N=100~\text{nm}$, $\alpha=2.5~\text{meV nm}$, $\mu=46.86~\text{meV}$, $\mu_J=0.8\mu$, $\mu_{SC}=\mu$, $m^*=0.036m_e$, where $m_e$ is the electron mass, and lattice constant for the tight-binding approximation of $a=5~\text{nm}$. To simulate the experimental system in Fig.~\ref{models}(c) and Fig.~\ref{xi_study}(a), we take $\Delta_{ind}=0.2~\text{meV}$ and for the small coherence length calculation in Fig.~\ref{xi_study}(b) we take $\Delta_{ind}=1.6~\text{meV}$. 
For the energy and frequency range studied in this work, we take the truncation for the Floquet coefficients and effective Hamiltonian as $M=5$ and $N=400$, respectively. The value of $M$ is determined by taking the lowest frequency studied, setting the correspondent energy to be of order $\hbar\omega\sim\Delta_{ind}$. The cutoff for the effective Hamiltonian $N$ sets the bandwidth  $\mathcal{W}\sim\Delta_{ind}$, resulting in localization range $\ell_m\sim 1$. Therefore, the choice of $M=5$ is sufficient to describe the lowest frequency states ($\omega \gtrsim 0.5 \Delta_{ind}$) in this work. For the study of Floquet physics at lower frequencies, the range $\ell_m$ would increase, and the truncation $M$ should increase accordingly.

\section{Conductance in a three-terminal system}
\label{appendix:3terminal}

We consider a three-terminal system composed of a normal (N) lead and superconductor SC1 at voltages $V_B$ and $V_J$, respectively, relative to a grounded superconductor SC2, as shown in Fig.~\ref{models}(b). The differential conductance in the normal lead $d I_N / d V_B$ will have contributions from two NS interfaces, each of which is described by a low-temperature tunneling conductance given by $dI_{NS}/dV=G_{NN}N_{BCS}(e|V|)/N_N$ \cite{tinkham2004introduction}, where $V$ is the voltage drop across the interface, $G_{NN}$ is the interface normal state conductance (assumed constant in $V$), and $N_{BCS}(e|V|)$ and $N_N$ are the Bardeen–Cooper–Schrieffer (BCS) superconducting and normal state DOS, respectively. 
The BCS DOS can be written with the Dynes formula \cite{PhysRevLett.41.1509,PhysRevB.94.144508}
\begin{equation}
    \label{eq:BCSDOS}
    N_{BCS}(\epsilon)=\operatorname{Re}\left[ \frac{\epsilon+i\gamma}{\sqrt{(\epsilon+i\gamma)^2-\Delta^2}}\right],
\end{equation}
where $\gamma$ is a phenomenological broadening parameter that quantifies the pair-breaking process in the tunneling density of states, and $\Delta$ is the superconducting gap. 

The superconducting gap is a function of the system parameters, such as the applied bias current $I_{DC}$ and magnetic field. At low supercurrents  $I_S$, the pairing potential is suppressed approximately as $\sqrt{1-(I_S/I_C)^2}$, where  $I_C$ is the bulk superconductor critical current~\cite{tinkham2004introduction}. Since Josephson voltage $V_J$ varies approximately linearly with current (see Fig.~\ref{main}(b)), we can write this suppression in terms of $V_J$. Similarly, the magnetic field will suppress the superconducting gap. Here, we neglect spin-orbit scattering effects and write a linear Pauli suppression due to the Zeeman field, which agrees well with the behavior found experimentally. All together, we can write the induced pairing potential suppressed by current and magnetic field as 
\begin{equation}
\label{eq:pairing_suppresed}
    \Delta=\Delta_{ind}\sqrt{1-\left( \frac{V_J}{V_C} \right)^2}\left(1-\frac{B}{B_C}\right),
\end{equation}
Here $\Delta_{ind}$ is the induced superconducting gap of InAs in the absence of a magnetic field and supercurrent, $V_C$ is the voltage drop across Josephson junction at the critical current, $B_C$ is the critical in-plane magnetic field. The gap suppression is fully characterized by two phenomenological parameters, $V_C$ and $B_C$, which can be obtained from the data. The parameter $V_C$ is obtained from the experimental data at $B=0~$T shown in Fig.~\ref{in-plane}(a) to match the voltage $V_J$ in which the superconducting gap closes at $V_J \approx 550~\mu$V. The value of $B_C$ can be found by extrapolating the data shown in Fig.~\ref{in-plane}(i) and finding the crossing point between P1 and P3, i.e., the Zeeman field at which the superconducting gap closes. For the simulation plots shown in Fig.~\ref{models}(d) and Fig.~\ref{in-plane}(e)-(h) and (j) we use $\Delta_{ind} = 0.2~$meV, $V_C=2.75\Delta_{ind}$ and $B_C=650~$mT. The conductance as a function of the voltages $V_J$ and $V_B$ is found by adding the two contributions from the NS interfaces, $d I_N / d V_B \propto N_{BCS}(e|V_J-V_B|)+N_{BCS}(e|V_B|)$, where the voltages are independent of the magnetic field in Figs.~\ref{in-plane}(e)-(h). However, for the conductance shown in Fig.~\ref{in-plane}(j), the experiment is carried out with fixed bias current $I_{DC}$, independent of the magnetic field. This is achieved by varying the bias voltage $V_J$ as a function of the applied field; see Fig.~\ref{V_j_vs_B}. The simulated conductance shown in Fig.~\ref{in-plane}(j) is obtained by taking $V_J\rightarrow V_J(B_\parallel)$ from experimental data, Fig.~\ref{V_j_vs_B}.

\section{wafer characterization}
\label{appendix:wafer characterization}

\begin{figure*}
\includegraphics[width=1\textwidth]{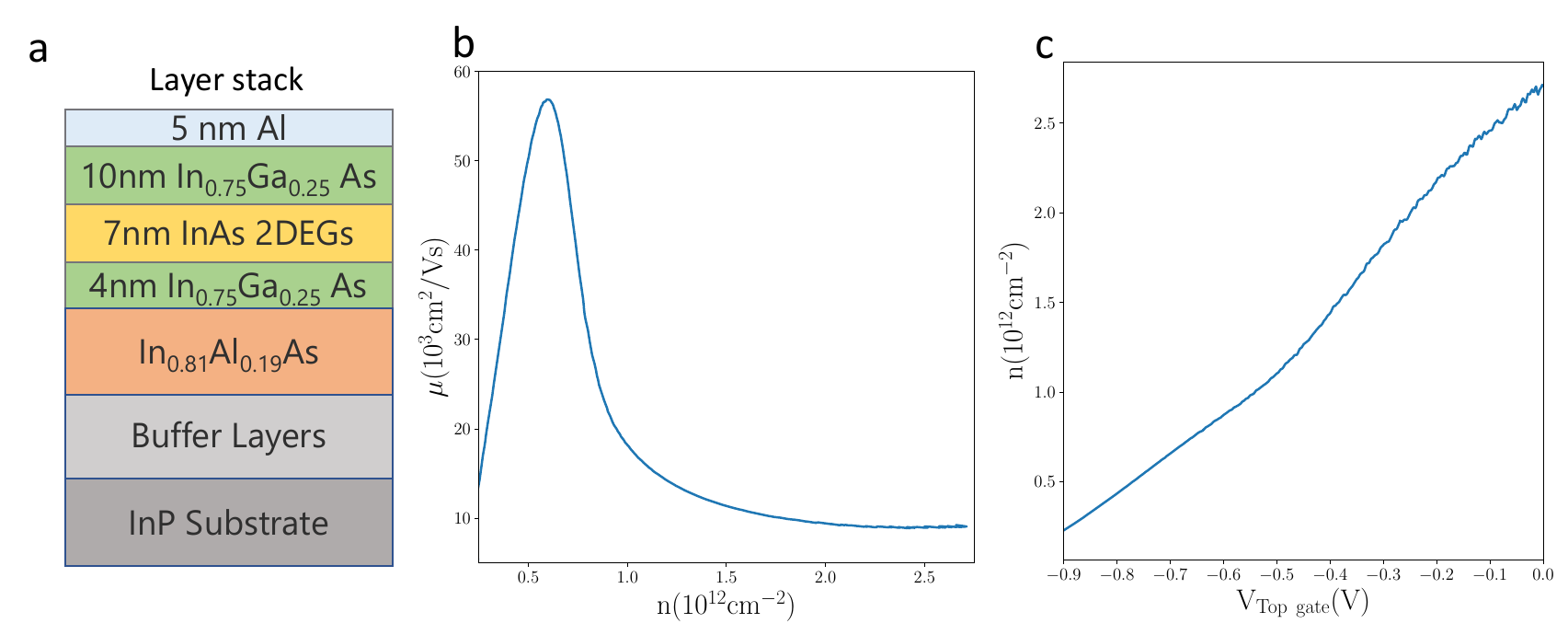}
\caption{\label{mobility_denisty}(a) Layer stack of heterostructure used in this work. (b) Mobility $\mu$ as a function of 2DEG density $n$ extracted from the gated hall bar measurements. (c) Density $n$ as a function of the top gate voltage in the hall bar $V_{\text{Top, gate}}$.}
\end{figure*}
 The heterostructure used in this work, shown schematically in Fig.~\ref{mobility_denisty}(a), was grown by molecular beam epitaxy. More growth and fabrication details can be found in Refs.~\cite{Zhang.2023, GARDNER201671}. To characterize basic 2DEG properties, gated hall bars were fabricated and measured in a cryogen-free dilution refrigerator at a mixing chamber temperature of $T=10$ mK. Additional details on basic characterization measurements can be found in~\cite{Zhang.2023}. In Fig.~\ref{mobility_denisty}(b), we plot the mobility $\mu$ as a function of the 2DEG density $n$, where we find the peak mobility $57\ 000$ cm$^2$ / V at $n= 0.6 \times 10^{12}$ cm$^{-2}$. In Fig.~\ref{mobility_denisty}(c), the 2DEG density is plotted as a function of the top gate voltage $V_{\text{Top, gate}}$; Peak mobility occurs at $V_{\text{Top, gate}} = -0.7~$V. 

\begin{figure}
\includegraphics[width=0.48\textwidth]{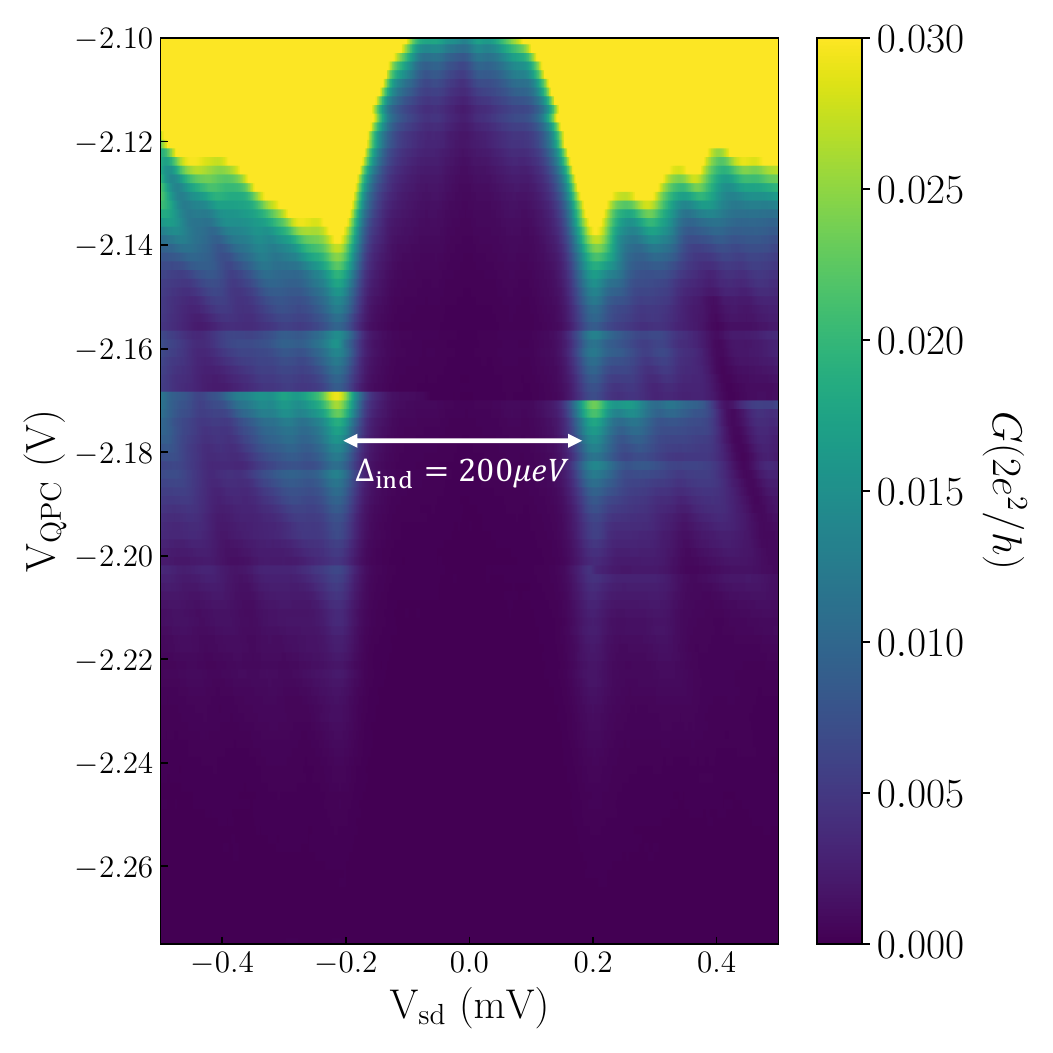}
\caption{\label{Induced gap} Tunneling conductance $G$ as a function of QPC gate voltage $V_{QPC}$ and source-drain bias $V_{\text{sd}}$. The white line indicates the magnitude of the induced gap, $\Delta_{ind} = 200~\mu $eV.}
\end{figure}

To investigate the induced superconducting gap in this wafer, we fabricated superconductor-quantum point contact-semiconductor (SQPCN) devices designed to perform tunneling spectroscopy in the epi-Al/InAs region. The quantum point contact gate voltage controls the tunneling barrier potential $V_{\text{QPC}}$. The differential conductance $G$ is measured as a function of the source-drain bias $V_{\text{sd}}$. When the QPC is tuned to the tunneling regime, the differential conductance reflects the local density of states (LDOS), allowing direct observation of the induced superconducting gap. In Fig.~\ref{Induced gap} we show tunneling spectroscopy as a function of the QPC gate voltage $V_{\text{QPC}}$. The magnitude of the induced gap, extracted from the separation of the coherence peaks, is approximately $200 \, \mu\text{eV}$. Fabrication and measurement techniques are consistent with those described in~\cite{Zhang.2023}.

\section{Interface Transparency}
\label{appendix:Transparency}
\begin{figure}
\includegraphics[width=0.4\textwidth]{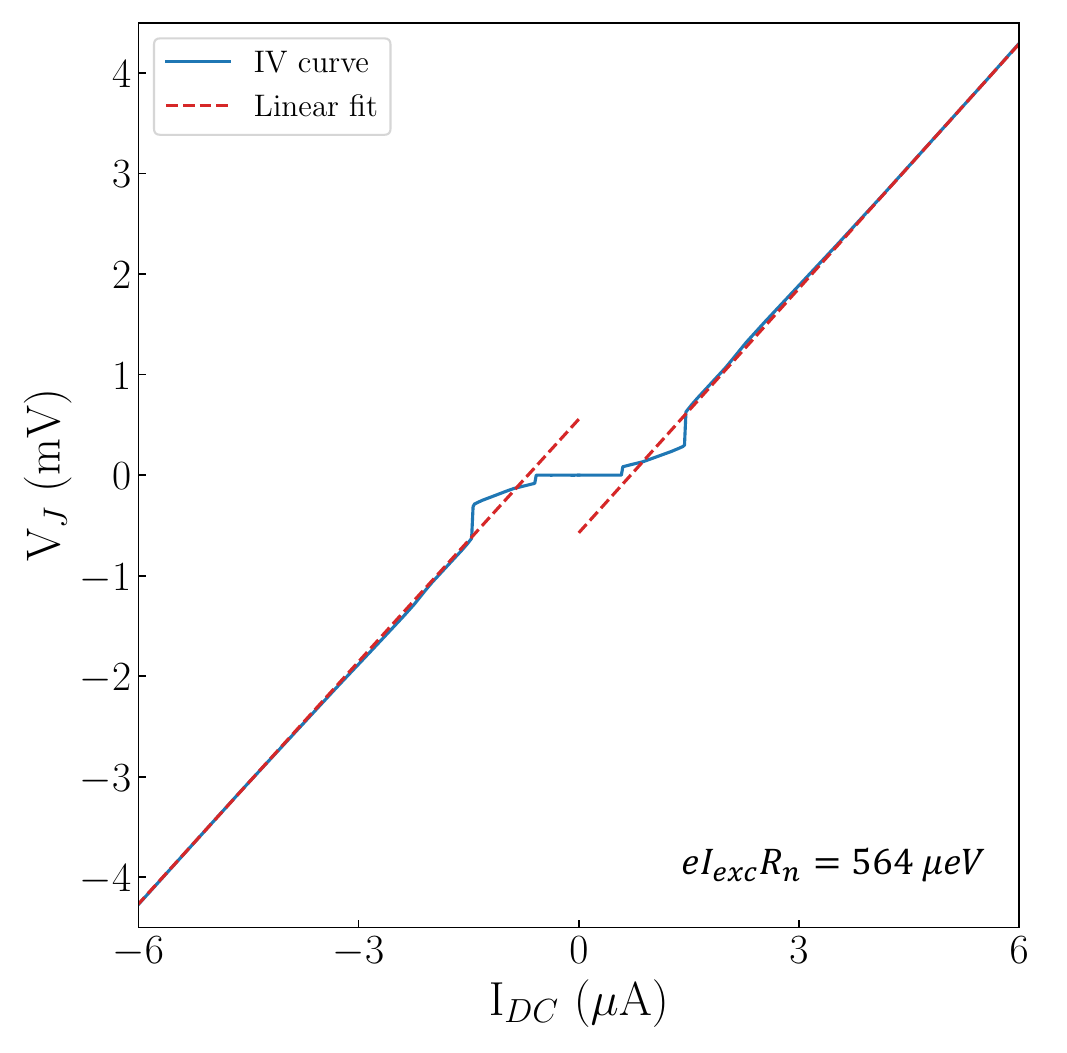}
\caption{\label{Trans} At $B_{\parallel}$=0~T and $V_{middle} = -0.7~V$, the blue line plots the junction voltage, $V_J$, as a function of DC current,$I_{DC}$, with $V_T=0~$V and $V_B = 0~$V. The red dashed line is a linear fit in the normal region ($|I_{DC}| \geq 4~\mu$A) used to extract the excess current, $I_{exc}$, and the normal state resistance, $R_N$. $eI_{exc}R_N$ extracted from this linear fit is $564 ~\mu e$V.}
\end{figure}

Fig.~\ref{Trans} displays I-V data (blue) at $V_{middle} = -0.7~V$ and a zero magnetic field on the same planar Josephson junction as studied in the main text. A linear fit is performed in the large current regime ($|I_{DC}| \geq 4~\mu\text{A}$) to extract excess current, $I_{exc}$, and the normal state resistance $R_N$. The intercept between the linear fit and the x-axis defines the excess current. The product of excess current and normal state resistance, $eI_{exc}R_N/\Delta_{ind}$, can be used to estimate the transparency of the interface between the bare InAs region and the proximitized epi-Al/InAs region \cite{Niebler.2009, Blonder.1982}. For a perfectly transparent interface, $eI_{exc}R_N/\Delta_{ind} \approx 2.7$ in the ballistic regime \cite{Niebler.2009, Blonder.1982}. From the linear fit, $eI_{\text{exc}}R_N$ is $564 ~\mu\text{eV}$, yielding $eI_{exc}R_N/\Delta_{ind} = 564 ~\mu\text{eV} / 200~\mu\text{eV} = 2.8$, which is close to the value of $2.7$ expected for a fully transparent interface. This indicates high transparency between the bare InAs and epi-Al/InAs regions of the studied Josephson junction.

\section{Reduction in Critical Current Associated with Normal Current Injection}
\label{appendix:Critical Current Decrease}
\begin{figure}
\includegraphics[width=0.4\textwidth]{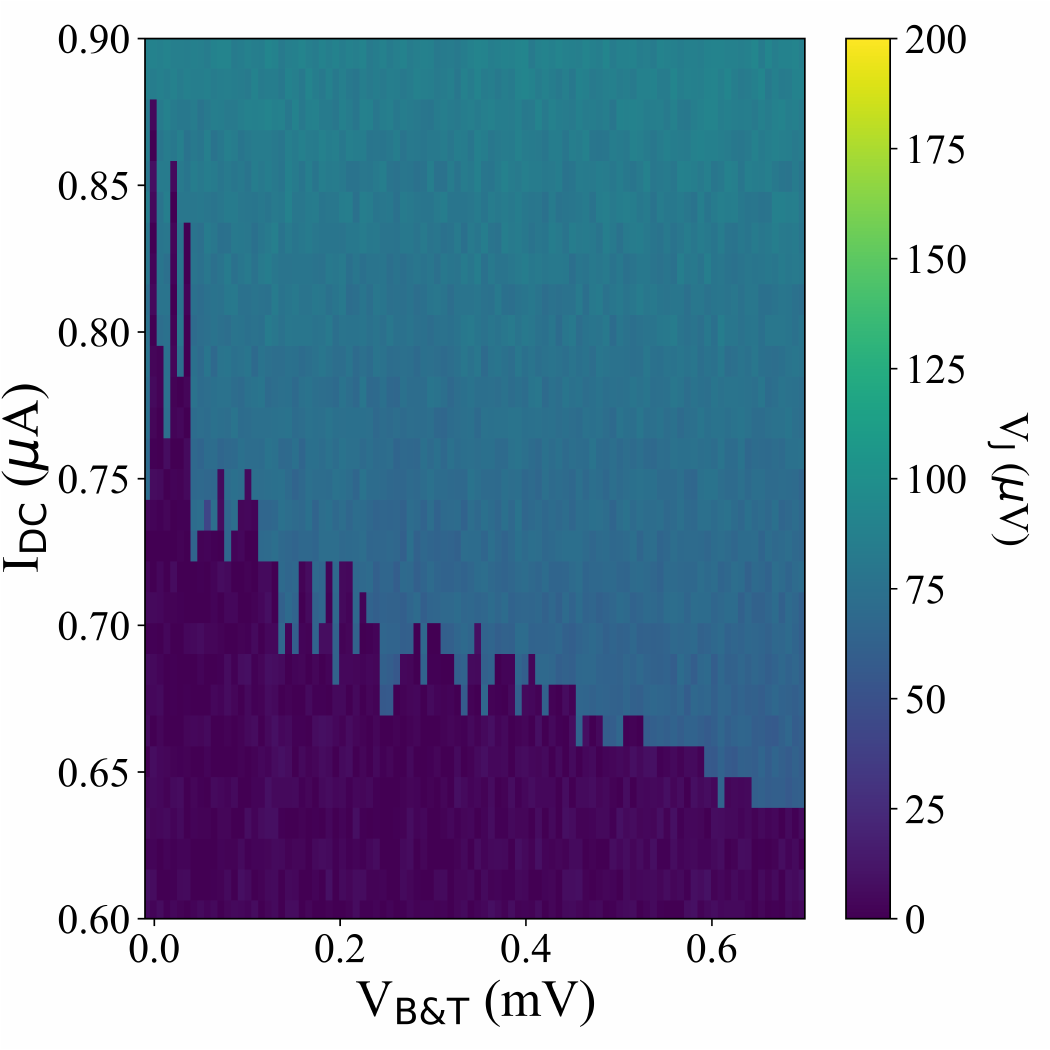}
\caption{\label{Ic_decrease} DC voltage drop across Josephson junction, $V_{J}$, as a function of DC voltage bias on both top and bottom tunnel probes, $V_{T(B)}$, and DC current bias $I_{DC}$ in a Josephson junction device with $W_s = 600~$nm, $L = 1.6~\mu$m, $W_N$ = 100~nm at the zero magnetic field. The dark blue region ($V_{J} = 0$) indicates the DC Josephson regime.}
\end{figure}
Fig.~\ref{Ic_decrease} presents the DC voltage drop across the Josephson junction, $V_{J}$, as a function of the DC voltage bias simultaneously applied to the top and bottom tunnel probes, $V_{T}$ and $V_{B}$, and the DC current bias, $I_{DC}$. This measurement was performed in a Josephson junction with a superconducting lead width $W_s = 600~$nm, junction length $L = 1.6~\mu$m, and normal region width $W_N = 100~$nm. The transmissions of the top and bottom quantum point contact gates were tuned to set the high-bias conductance to $\sim0.05G_0$, where $G_0$ is the quantum conductance. Fig.~\ref{Ic_decrease} shows a decrease in the critical current from $0.88$$~\mu$A at $V_{T(B)}) = 0~$V to $0.64~\mu$A at $V_{T(B)} = 0.7~$mV. In Ref.~\cite{Schäpers.1998r3g}, a similar reduction in critical current is reported due to electron injection from an additional 2DEG lead in ballistic Nb-2DEG-Nb Josephson junctions, similar to our configuration. In our devices, tunneling currents are injected from the top and bottom tunneling probes, which behave similarly to the additional 2DEG lead in Ref.~\cite{Schäpers.1998r3g}. The authors of Ref.~\cite{Schäpers.1998r3g} suggest two mechanisms for this phenomenon: 1) coupling between the Josephson junction and additional normal contacts (our tunneling probes) leads to quasiparticle phase breaking and suppression of the net critical current, 2) electrons injected by DC voltage-biased contacts occupy previously empty Andreev bound states, thereby modulating the net critical current.

\section{Measurements of Josephson Junctions in Various Geometries}
\label{appendix:geometry}
\begin{figure*}
\includegraphics[width=0.8\textwidth]{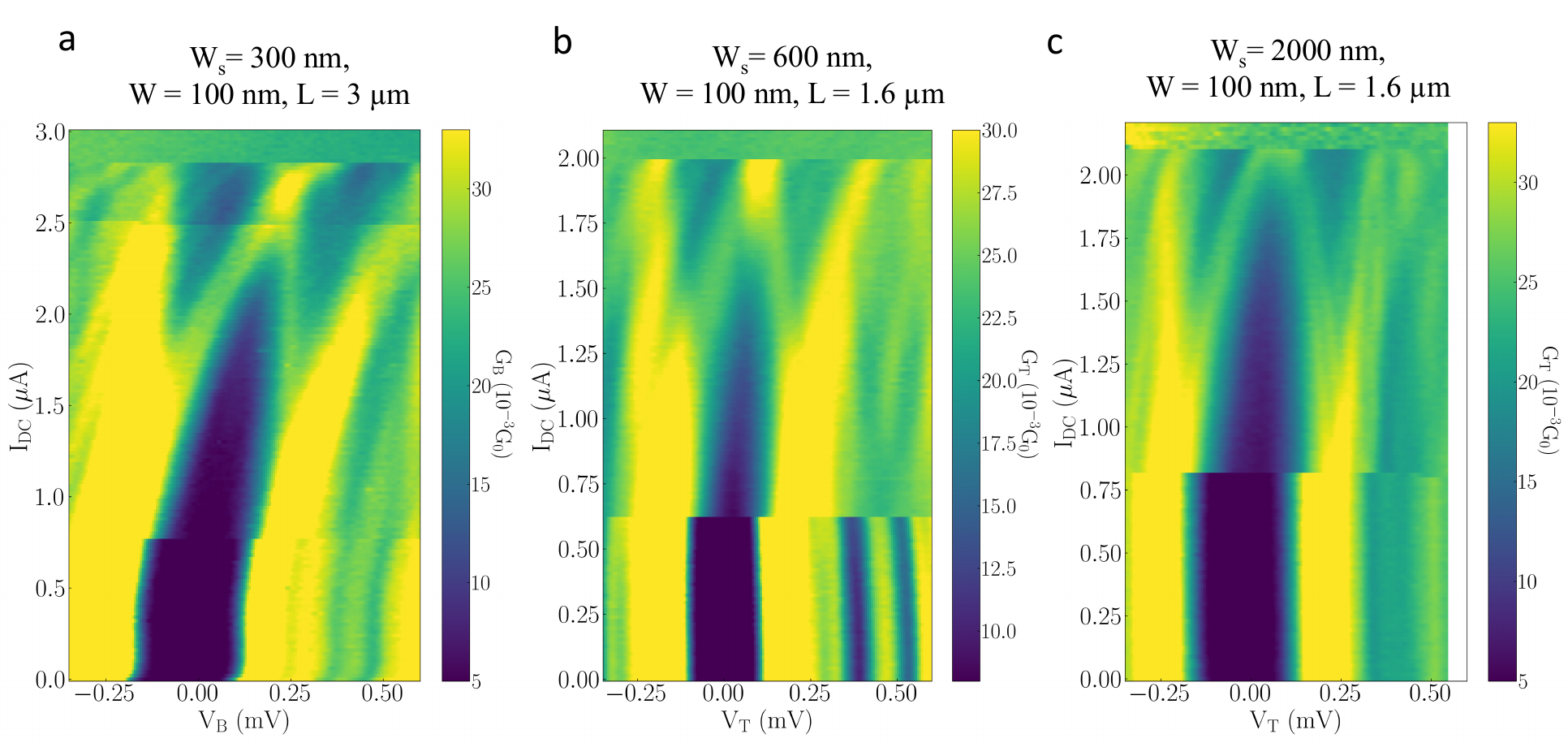}
\caption{\label{geometry_study} (a) - (c): Tunneling spectra as a function of DC current bias at zero in-plane magnetic field for Josephson junctions in different geometries. Geometry details for each device are indicated above each plot.}
\end{figure*}

The spectral features in the AC Josephson regime discussed in the main text were observed in four Josephson junctions with different geometries, as shown in Fig.~\ref{geometry_study}. These data indicate that the phenomenon is robust and is not sensitive to the width of the superconducting leads, $W_S$, or the length of the junction, $L$, within the ranges studied. Note that the middle gate voltages, $V_{middle}$, applied in Fig.~\ref{geometry_study}(a)-(c) vary: $V_{middle} = -0.7~V$ in (a), $V_{middle} = -0.39~V$ in (b) and $V_{middle} = 0~V$ in (c).

\section{Polarity of Applied DC Current}
\label{appendix:Bias Current directions}
\begin{figure}
\includegraphics[width=0.47\textwidth]{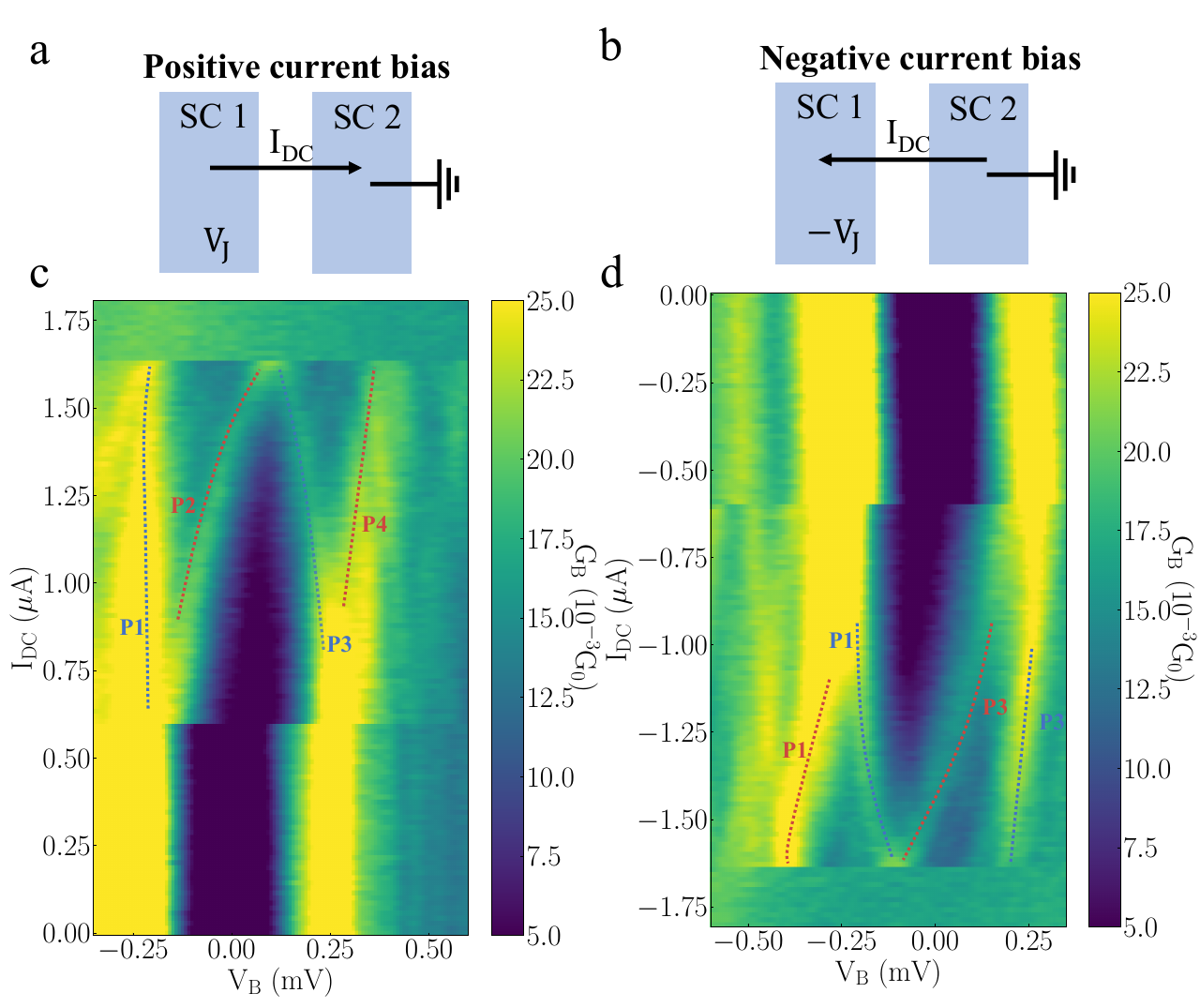}
\caption{\label{negative_IDC} (a) and (b): Schematic of positive current bias and negative current bias. (c) and (d): Tunneling spectra as a function of the DC current bias at zero in-plane magnetic field for positive and negative bias on the same Josephson junction reported in the main text. Conductance peaks in (c) and (d) are indicated by dashed lines.}
\end{figure}

We also tested for the polarity dependence of the DC current bias. Fig.~\ref{negative_IDC} (a) and (b) are schematics of the junction with positive and negative current biases, respectively; the main difference is that the potential of the left superconducting lead changes to $-V_J$ in the negative current bias case. Fig.~\ref{negative_IDC}(c) and (d) display the tunneling spectra as a function of the DC current bias, $I_{DC}$, using different current bias directions at zero magnetic field on the same devices studied in the main text. Note that the middle gate voltage is set at -0.5 V during the measurements of Fig.~\ref{negative_IDC} (c) and (d). Conductance peaks P1 and P3 are centered around zero voltage for both positive and negative current bias cases. However, the centers of conductance peaks P2 and P4 shift to positive voltage in the positive current bias case and to negative voltage in the negative case, as indicated by dashed lines in Fig.~\ref{negative_IDC}(c) and (d). This behavior is in agreement with the predictions of the three-terminal model.

\section{$V_J$ as a function of $B_{\parallel}$ at $I_{DC}$ = 1.1~$\mu$A}
\label{appendix:vJ vs B}

\begin{figure}
\includegraphics[width=0.4\textwidth]{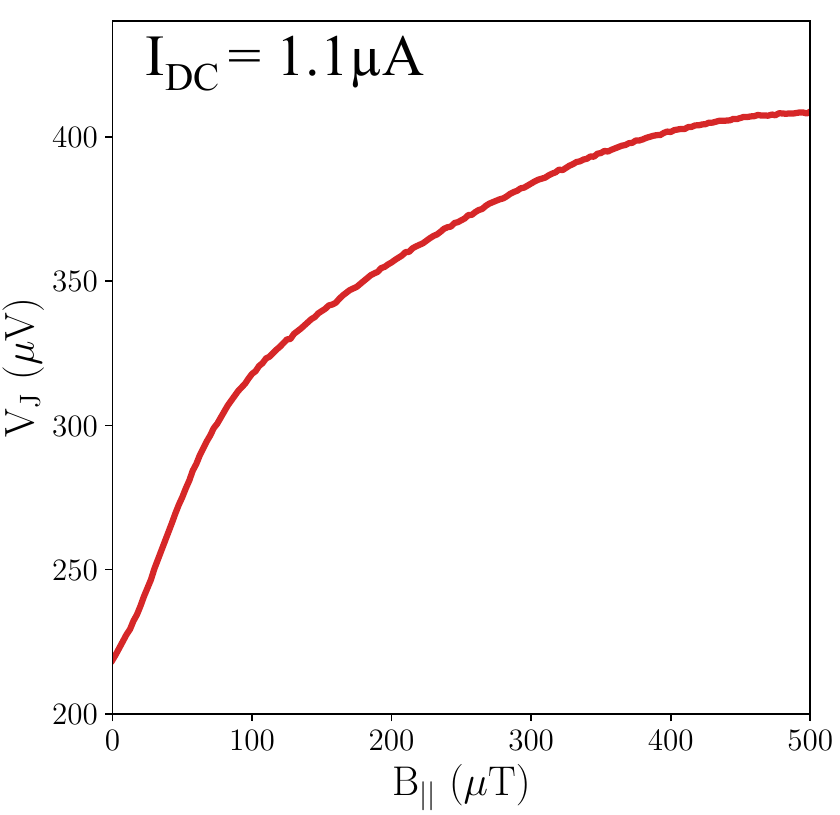}
\caption{\label{V_j_vs_B} At $I_{DC} = 1.1~\mu$A, $V_J$ as a function of $B_{||}$, measured simultaneously with the data shown in Fig.~\ref{in-plane}(i).}
\end{figure}

Fig.~\ref{V_j_vs_B} displays the DC voltage drop, $V_J$, as a function of the in-plane magnetic field, $B_{||}$ measured simultaneously with the data shown in Fig.~\ref{in-plane}(i). It is evident that $V_J$ increases with $B_{||}$, explaining the observed shift of the center of P2 and P4 to higher voltages as $B_{||}$ increases in Fig.~\ref{in-plane}(i). This increase in $V_J$ at a constant DC current bias has been previously documented in the literature \cite{Lee.2019, Dartiailh.2021, Suominen.2017}. In the AC Josephson regime, the resistance primarily arises from single quasiparticle transport across the junction, mediated by multi-Andreev reflections (MARs). We observed MARs feature in our devices (not shown). This process is influenced by the in-plane magnetic field, leading to the observed behavior.
\end{document}